\documentclass[amsmath,amssymb,superscriptaddress,nobalancelastpage,prb,twocolumn,showpacs]{revtex4}
\usepackage{graphicx}
\usepackage{color}

\begin{document}

%%%%%%%%%%%%%%%%%%%%%%%%%%%% TITLE

\title{Nernst effect in the cuprate superconductor YBa$_2$Cu$_3$O$_y$: Broken rotational and translational symmetries}

%%%%%%%%%%%%%%%%%%%%%%%%%%%% AUTHORS

\author{J. Chang}
\affiliation{D\'epartement de physique
\& RQMP, Universit\'e de Sherbrooke, Sherbrooke, Canada}

\author{Nicolas Doiron-Leyraud} 
\affiliation{D\'epartement de physique
\& RQMP, Universit\'e de Sherbrooke, Sherbrooke, Canada}

\author{Francis Lalibert\'e} 
\affiliation{D\'epartement de physique
\& RQMP, Universit\'e de Sherbrooke, Sherbrooke, Canada}

\author{R. Daou}
\altaffiliation{Present address: Max Planck Institute for Chemical Physics of Solids, Dresden 01187, Germany.}
\affiliation{D\'epartement de physique
\& RQMP, Universit\'e de Sherbrooke, Sherbrooke, Canada}

\author{David LeBoeuf}
\altaffiliation{Present address: Laboratoire National des Champs Magn\'etiques Intenses, UPR 3228 (CNRS, INSA, UJF, UPS), Toulouse 31400, France.}
\affiliation{D\'epartement de physique
\& RQMP, Universit\'e de Sherbrooke, Sherbrooke, Canada}

\author{B.~J. Ramshaw} 
\affiliation{Department of Physics \& Astronomy,
University of British Columbia, Vancouver, Canada}

\author{Ruixing Liang} 
\affiliation{Department of Physics \& Astronomy,
University of British Columbia, Vancouver, Canada}
\affiliation{Canadian Institute for Advanced Research, Toronto, Canada}

\author{D.~A. Bonn} 
\affiliation{Department of Physics \& Astronomy,
University of British Columbia, Vancouver, Canada}
\affiliation{Canadian Institute for Advanced Research, Toronto, Canada}

\author{W.~N. Hardy} 
\affiliation{Department of Physics \& Astronomy,
University of British Columbia, Vancouver, Canada}
\affiliation{Canadian Institute for Advanced Research, Toronto, Canada}

\author{Cyril Proust}
\affiliation{Laboratoire National des Champs Magn\'etiques Intenses, UPR 3228 (CNRS, INSA, UJF, UPS), 31400 Toulouse, France}
\affiliation{Canadian Institute for Advanced Research, Toronto, Canada}

\author{I. Sheikin}
\affiliation{Laboratoire National des Champs Magn\'etiques Intenses (CNRS), 38042 Grenoble, France}

\author{K. Behnia}
\affiliation{LPEM (UPMC-CNRS), ESPCI, 75231 Paris, France}

\author{Louis Taillefer}
\altaffiliation{E-mail: louis.taillefer@physique.usherbrooke.ca.}
\affiliation{D\'epartement de physique 
\& RQMP, Universit\'e de Sherbrooke, Sherbrooke, Canada} 
\affiliation{Canadian Institute for Advanced Research, Toronto, Canada}

\date{\today}

%%%%%%%%%%%%%%%%%%%%%%%%%%%% ABSTRACT

\begin{abstract}

The Nernst coefficient of the cuprate superconductor YBa$_2$Cu$_3$O$_{y}$ was recently shown to become strongly anisotropic 
within the basal plane when cooled below the pseudogap temperature $T^\star$, revealing that the
pseudogap phase breaks the four-fold rotational symmetry of the CuO$_2$ planes.
Here we report on the evolution of this Nernst anisotropy at low temperature, once superconductivity is suppressed 
by a magnetic field. 
We find that the anisotropy drops rapidly below 80 K, to vanish in the $T=0$ limit.
We show that this loss of anisotropy is due to the emergence of a small high-mobility electron-like pocket in the Fermi surface 
at low temperature, a reconstruction attributed to a low-temperature state that breaks the translational symmetry of the CuO$_2$ planes.
We discuss the sequence of broken symmetries -- first rotational, then translational -- in terms of an electronic nematic-to-smectic transition 
such as could arise when unidirectional spin or charge modulations order.
We compare YBa$_2$Cu$_3$O$_{y}$ with iron-pnictide superconductors where the process of (unidirectional) antiferromagnetic ordering 
gives rises to the same sequence of broken symmetries.

\end{abstract}

%\pacs{74.72.Dn, 78.70.Nx, 74.25.Nf}

%74.25.Fy Transport properties (electric and thermal conductivity, thermoelectric effects, etc.)
%74.72.Dn %La-based cuprates

\maketitle

%%%%%%%%%%%%%%%%%%%%%%%%%%%% INTRODUCTION
\section{Introduction}
%%%%%%%%%%%%%%%%%%%%%% FIGURE 1 %%%%%%%%%%%%%%%%%%%%%%%%%%%%%%%%%%%%

%%%%%%%%%%%%%%%%%%%%%%%%%%%%%%%%%%%%%%%%%%%%%%%%%%%%%%%%%%%%%%%%%%%%
Establishing and understanding the normal-state 
phase diagram of cuprates is of primary importance
in the quest to uncover the mechanism of 
high-temperature superconductivity.
The discovery of quantum oscillations in underdoped YBa$_2$Cu$_3$O$_{y}$ (YBCO) revealed 
that in the absence of superconductivity, suppressed by application of a large magnetic field,
the ground state in the underdoped region of the phase diagram is a metal whose Fermi surface contains a small closed 
pocket.\cite{Doiron-Leyraud2007}
The negative Hall and Seebeck coefficients of YBCO at $T \to 0$ show this pocket to be electron-like.\cite{LeBoeuf2007,Chang2010}
The presence of an electron pocket in the Fermi surface of a hole-doped cuprate is the typical signature 
of a Fermi-surface reconstruction caused by the onset of a new periodicity which breaks the
translational symmetry of the crystal lattice.\cite{Taillefer2009,Chakravarty2008}
%An example is antiferromagnetic, or spin-density-wave (SDW), order.\cite{Harrison2009}  
%
\begin{figure}[h!]
\begin{center}
\includegraphics[width=0.375\textwidth]{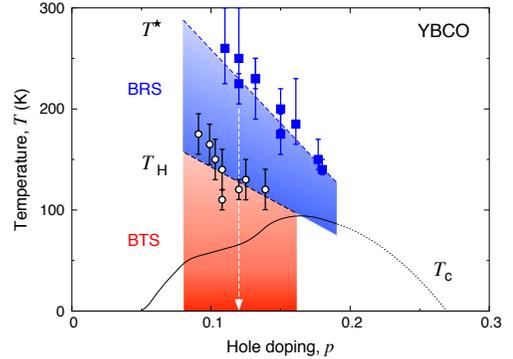}
\caption{
Phase diagram of YBCO, showing the zero-field superconducting transition temperature $T_c$
(dotted line, Ref.~\onlinecite{Liang2006}, extrapolated as dashed line above $p=0.18$) 
and the pseudogap crossover temperature $T^\star$ detected by the 
Nernst effect (squares; Ref.~\onlinecite{Daou2010}).
The onset of in-plane anisotropy in the Nernst coefficient below $T^\star$ shows that the pseudogap phase
is a state with broken rotational symmetry (BRS).\cite{Daou2010}
Once superconductivity is suppressed by a magnetic field, the normal state at $T \to 0$ is characterized 
by a reconstructed Fermi surface,\cite{Doiron-Leyraud2007,LeBoeuf2007}
evidence of broken translational symmetry (BTS).
The temperature $T_{\rm H}$ below which the Hall coefficient $R_{\rm H}(T)$ starts to deviate downward is the first signature 
of Fermi-surface reconstruction upon cooling (open circles; from Ref.~\onlinecite{LeBoeuf2011}).
The white down-pointing arrow locates the present study of Nernst anisotropy on the phase diagram.
The two dashed lines are a guide to the eye.
}
\label{fig:Phasediagram}
\end{center}
\end{figure}

In the doping phase diagram of YBCO (Fig.~1), the electron pocket exists at $T \to 0$ (in the absence of superconductivity) 
throughout the range from $p = 0.083$ to at least $p = 0.152$.\cite{LeBoeuf2011} 
We infer that translational symmetry is broken at $T = 0$ over at least that range,
by an ordered phase that has yet to be definitively identified. 
The fact that the Seebeck coefficient of the cuprate La$_{1.8-x}$Sr$_x$Eu$_{0.2}$CuO$_4$ (Eu-LSCO) at $T \to 0$ 
is negative over the same range of doping as in YBCO,\cite{Laliberte2011}
and that stripe order -- a unidirectional modulation of spin and/or charge densities\cite{Kivelson2003,Vojta2009} -- 
prevails in Eu-LSCO over that doping range,\cite{Fink2010}
is compelling evidence that stripe order is responsible for the broken translational symmetry 
and that Fermi-surface reconstruction is a generic property of hole-doped cuprates.

%%%%%%%%%%%%%%%%%%%%%% FIGURE 2 %%%%%%%%%%%%%%%%%%%%%%%%%%%%%%%%%%%%

\begin{figure}
%\begin{center}
\includegraphics[width=0.4\textwidth]{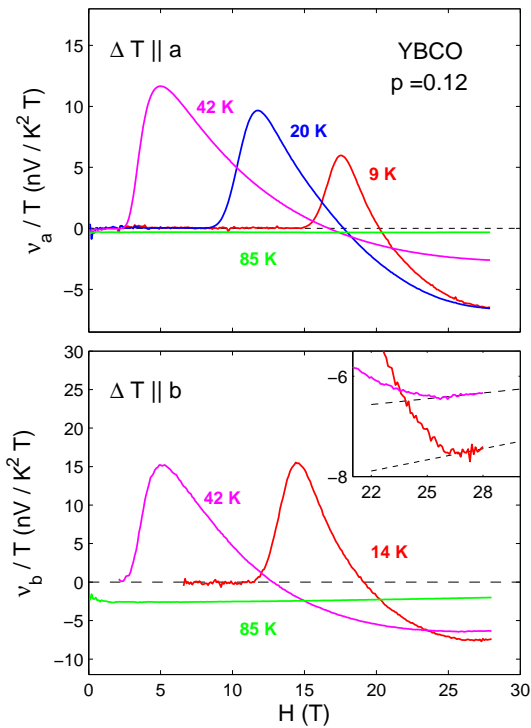}
\caption{
Nernst coefficient $\nu$ of YBCO at a hole concentration (doping) $p=0.12$, 
plotted as $\nu/T$ versus magnetic field $H$,
for different temperatures, as indicated. 
Top: the temperature gradient $\Delta T$ is applied along the $a$ axis of the orthorhombic crystal structure.
Bottom: $\Delta T$ is along the $b$ axis. 
Inset:
zoom on the data at high field. The saturation in $\nu$ vs $H$ above $H \simeq 26$~T
indicates that the positive contribution from superconducting fluctuations has become negligible, and 
the data above 26 T represent the normal-state properties of YBCO at that doping.
}
\label{fig:nuvsH}
%\end{center}
\end{figure}

%%%%%%%%%%%%%%%%%%%%%%%%%%%%%%%%%%%%%%%%%%%%%%%%%%%%%%%%%%%%%%%%%%%%

At temperatures above the superconducting transition temperature $T_c$,
the normal-state phase diagram of cuprates is characterized by the pseudogap phase,
below a crossover temperature $T^\star$.\cite{Norman2005} 
The Nernst effect was recently found to be a sensitive probe of the pseudogap phase,\cite{Cyr-Choiniere2009,Matusiak2009,Daou2010}
such that it can be used to detect $T^\star$, as shown in Fig.~1 for YBCO.
Measurements of the Nernst coefficient $\nu(T)$ in detwinned crystals of YBCO for a temperature gradient along the $a$-axis 
and $b$-axis directions within the basal plane of the orthorhombic crystal structure revealed a strong in-plane anisotropy, 
setting in at $T^\star$.\cite{Daou2010} 
This showed that the pseudogap phase breaks the four-fold rotational symmetry of the CuO$_2$ planes, 
throughout the doping range investigated, from $p = 0.08$ to $p = 0.18$.\cite{Daou2010}

In this paper, we investigate the impact of Fermi-surface reconstruction on this Nernst anisotropy, 
by extending the previous Nernst study to low temperature.
We find that below 80 K the anisotropy falls rapidly, in close parallel with the fall in the Hall coefficient to negative values.  
We infer that the Nernst anisotropy disappears because the small closed high-mobility electron pocket which dominates the
transport properties of YBCO at low temperature yields isotropic transport.

%%%%%%%%%%%%%%%%%%%%%% FIGURE 3 %%%%%%%%%%%%%%%%%%%%%%%%%%%%%%%%%%%%

\begin{figure}
%\begin{center}
\includegraphics[width=0.4\textwidth]{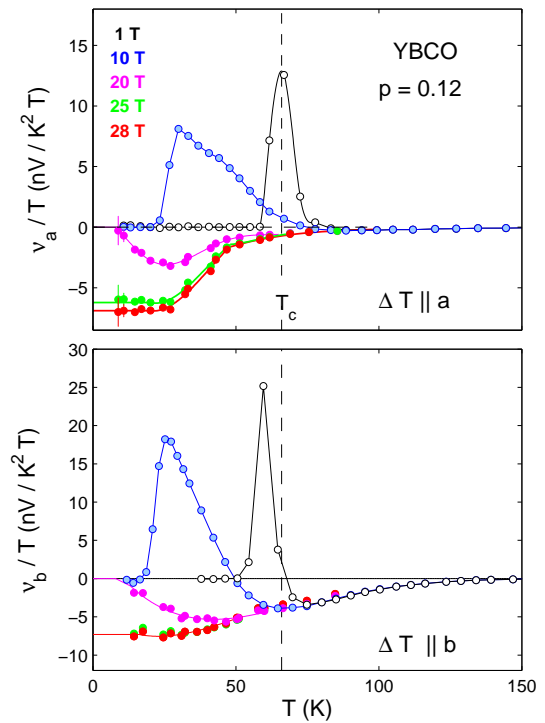}
\caption{
Nernst coefficient $\nu$ of YBCO at $p=0.12$, 
plotted as $\nu/T$ versus temperature $T$,
for different values of the magnetic field, as indicated. 
Top: $\Delta T$ is along the $a$ axis.
Bottom: $\Delta T$ is along the $b$ axis. 
The vertical dashed line marks the zero-field superconducting transition,
at $T_c = 66.0$~K.}
\label{fig:nuvsT}
%\end{center}
\end{figure}

%%%%%%%%%%%%%%%%%%%%%%%%%%%%%%%%%%%%%%%%%%%%%%%%%%%%%%%%%%%%%%%%%%%%

%%%%%%%%%%%%%%%%%%%%%%%%%%%%%%%%%%%%%%%%%%%%%%%%%%%%%%%%%%%%%%%%%%%%
%%%%%%%%%%%%%%%%%%%%%%%%%%%%%% EXPERIMENTAL %%%%%%%%%%%%%%%%%%%%%%%%
%%%%%%%%%%%%%%%%%%%%%%%%%%%%%%%%%%%%%%%%%%%%%%%%%%%%%%%%%%%%%%%%%%%%

\section{Methods}
Measurements were performed on high-quality detwinned YBCO crystals
grown in a non-reactive BaZrO$_3$ crucible from 
high-purity starting materials.\cite{Liang2000}
The oxygen content was set at $y = 6.67$ and the 
dopant oxygen atoms were made to order into an 
ortho-VIII superstructure, yielding a superconducting
transition temperature $T_c = 66.0$~K. 
The hole concentration (doping) $p = 0.12$ was 
determined from a relationship between $T_c$ and 
the $c$-axis lattice constant \cite{Liang2006}. 
% 

%%%%%%%%%%%%%%%%%%%%%%  FIGURE 4 %%%%%%%%%%%%%%%%%%%%%%%%%%%%%%%%%%%

\begin{figure}
\begin{center}
\includegraphics[width=0.4\textwidth]{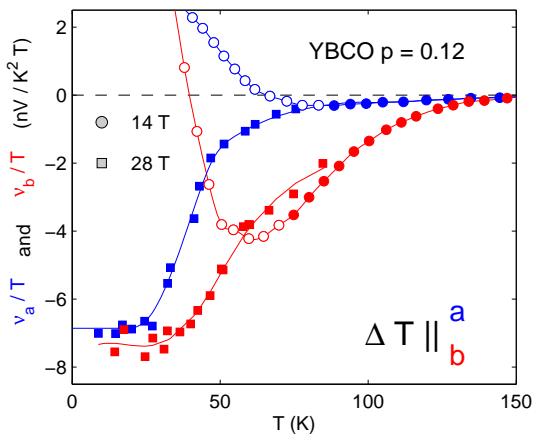}
\caption{
Nernst coefficient $\nu$ of YBCO plotted as $\nu/T$ vs $T$ for 
the two directions of temperature gradient 
($\nu_a$ for $\Delta T \parallel a$, blue symbols; $\nu_b$ for $\Delta T \parallel b$; red symbols), 
for two values of the magnetic field: $H=14$~T (circles) and 28~T (squares). 
Full symbols correspond to normal-state data, in which the superconducting contribution to the 
Nernst signal is negligible.
Note that the normal-state data for $\nu_b$ at 14 T and 28 T do not quite coincide because 
of a slight field dependence of $\nu_b$, akin to magnetoresistance (see isotherm at 85 K in Fig.~2). 
}  
\label{fig:nunbvsT}
\end{center}
\end{figure}

%%%%%%%%%%%%%%%%%%%%%%%%%%%%%%%%%%%%%%%%%%%%%%%%%%%%%%%%%%%%%%%%%%%%

%%%%%%%%%%%%%%%%%%%%%  FIGURE 5  %%%%%%%%%%%%%%%%%%%%%%%%%%%%%%%%%%%

\begin{figure}[h!]
\begin{center}
\includegraphics[width=0.45\textwidth]{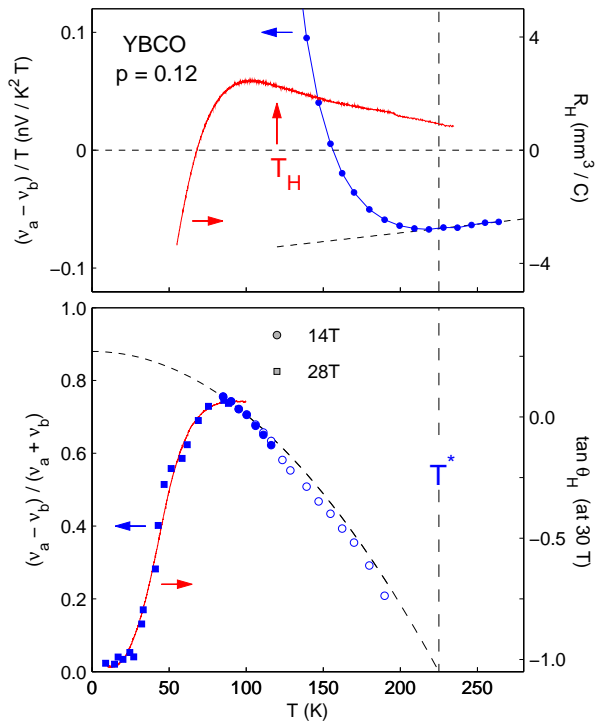}
\caption{
Temperature dependence of the in-plane anisotropy in the normal-state Nernst coefficient 
of YBCO at $p = 0.12$.
Top: 
Nernst anisotropy difference $(\nu_a - \nu_b)/T$ vs $T$
(blue circles, left axis; from Ref.~\onlinecite{Daou2010}).
The difference starts to rise below the pseudogap temperature $T^\star$ (vertical dashed line).
Also shown is the Hall coefficient $R_{\rm H}(T)$, measured in a field $H = 10$~T
(continuous red curve, right axis). 
Below $\sim 100$~K, $R_{\rm H}(T)$ drops precipitously to reach large negative values
at $T \to 0$.\cite{LeBoeuf2007,LeBoeuf2011} 
We can define the onset of this drop as $T_{\rm H}$, the temperature below which 
$R_{\rm H}(T)$ acquires downward curvature (see Ref.~\onlinecite{LeBoeuf2011}).
Bottom: 
Normalized Nernst anisotropy defined as the dimensionless ratio 
$\nu_a - \nu_b)/(\nu_a + \nu_b)$, 
plotted vs $T$ below 120~K (full blue symbols; squares, $H = 28$~T, this work;
circles, $H = 14$~T, from Ref.~\onlinecite{Daou2010}).
(Because both $\nu_a(T)$ and $\nu_b(T)$ cross zero around 150~K, the ratio becomes ill-defined above 120~K.
To avoid this, we define the sum and difference relative to their value at $T^\star$ and plot
their ratio above 120~K (open blue circles); see Ref.~\onlinecite{Daou2010}.)
Below $\sim 80$~K, the anisotropy ratio drops rapidly to zero as $T \to 0$ in a way that precisely tracks 
the drop in the Hall coefficient, shown here as tan$\theta_{\rm H}(T) \equiv \rho_{xy} / \rho_{xx}$ 
vs $T$ (continuous red curve, right axis), where the transverse ($\rho_{xy}$) 
and longitudinal ($\rho_{xx}$) resistivities are measured in 30~T.
This reveals that the disappearance of the Nernst anisotropy at low temperature is due to the Fermi-surface reconstruction
that leads to the formation of a small closed electron pocket of high mobility (see text).
}
\label{fig:NvsRH}
\end{center}
\end{figure}

%%%%%%%%%%%%%%%%%%%%%%%%%%%%%%%%%%%%%%%%%%%%%%%%%%%%%%%%%%%%%%%%%%%%%%%%%%%%%%%%%%

The Nernst effect, being the transverse voltage $V$ generated 
by a longitudinal temperature difference $\Delta T$ in 
a perpendicular applied magnetic field $H$,\cite{Ong2006,Behnia2009}
was measured in Sherbrooke up to 15 T and at the LNCMI in Grenoble up to 28 T.
In both cases, we used a 
one-heater two-thermometer setup and the field was 
applied along the $c$ axis of the orthorhombic crystal structure.
The Nernst signal was measured with the thermal gradient $\Delta T$ either along 
the $a$ axis ($\Delta T_a$) or the $b$ axis ($\Delta T_b$), 
and the Nernst coefficient $\nu$ is indexed as follows:
\begin{equation}
\nu_{a}=\frac{\alpha}{H}\frac{V_b}{\Delta T_a } \quad \textrm{and} \quad \nu_{b}=\frac{\alpha}{H}\frac{V_a}{\Delta T_b }
\end{equation}
where $\alpha=\ell/w$ is the ratio of sample length (between thermometer contacts) to
sample width.

%%%%%%%%%%%%%%%%%%%%%%%%%%%%%%%%%%%%%%%%%%%%%%%%%%%%%%%%%%%%%%%%%%%%
%%%%%%%%%%%%%%%%%%%% RESULTS %%%%%%%%%%%%%%%%%%%%%%%%%%%%%%%%%%%%%%%
%%%%%%%%%%%%%%%%%%%%%%%%%%%%%%%%%%%%%%%%%%%%%%%%%%%%%%%%%%%%%%%%%%%%

\section{Results}

Before we present our results, it is important to emphasize that there are two different contributions to the Nernst effect in a superconductor:
1) a positive contribution from superconductivity (moving vortices and fluctuations of the superconducting order parameter);
2) a contribution from normal-state quasi-particles, which can be either positive or negative. 
In YBCO, the two contributions can be readily separated because the quasi-particle contribution is negative, of opposite sign to the signal from superconducting 
fluctuations.\cite{Daou2010} 
Note also that the quasi-particle contribution to the Nernst coefficient $\nu(H)$ is mostly independent of field, 
whereas the superconducting contribution is strongly dependent on field.\cite{Ong2006}
In the electron-doped cuprate Pr$_{2-x}$Ce$_x$CuO$_4$, for example, this difference in the field dependence was used to separate the two contributions, 
both positive in this case.\cite{Li2007}

The amplitude of the quasi-particle contribution may be estimated from the following expression:\cite{Behnia2009} 
\begin{equation}
|\frac{\nu}{T}| \simeq \frac{\pi^2}{3} \frac{k_{\rm B}}{e} \frac{\mu}{T_{\rm F}}
\end{equation} 
where $k_{\rm B}$ is Boltzmann's constant, $e$ is the electron charge, 
$\mu$ is the carrier mobility and $T_{\rm F}$ the Fermi temperature. 
This relation, applicable in the  $T \to 0$ limit, 
was found valid within a factor of two for a wide range of metals.\cite{Behnia2009}
Its implication is that the Nernst effect is highly sensitive to Fermi-surface reconstructions 
that produce pockets with a small Fermi energy ($\epsilon_{\rm F} \equiv k_{\rm B} T_{\rm F}$) 
and a high mobility.
A good example of this is the heavy-fermion metal URu$_2$Si$_2$ where, upon cooling
below $17$~K, $\epsilon_{\rm F}$ drops by a factor of ten simultaneously with 
a ten-fold rise in the mobility $\mu$. 
As a consequence, $\nu/T$ rises by two orders of magnitude.\cite{Bel2004}

In Figs. 2 and 3, the Nernst coefficient $\nu$ of YBCO at $p = 0.12$
is plotted as $\nu/T$ vs $H$ and vs $T$, respectively, 
for both $a$- and $b$-axis directions. 
The high-field $b$-axis data are presented here for the first time,
while the low-field data\cite{Daou2010} and the high-field
$a$-axis data\cite{Chang2010} were reported previously.
We start by examining the isotherms (Fig.~2).
At $T < T_c = 66$~K, $\nu(H)$ shows the strong field dependence typical
of a superconductor:
1) at low field, $\nu = 0$ in the vortex solid phase;
2) at intermediate fields, $\nu$ rises to give a strong positive peak;
3) at higher field, the positive signal gradually decreases, until
such fields as $\nu(H)$ becomes flat, where the superconducting contribution 
has become negligible.
At the highest field measured in our experiment, 28 T, this saturation has been reached 
for all temperatures down to $\sim 10$~K, so that we may regard the state at 28 T (and above)
as the normal state.

At $T = 85$~K, $\nu_a(H)$ is seen to be totally flat and $\nu_b(H)$
increases very slightly (a form of normal-state magneto-resistance).
The positive (field-decreasing) superconducting contribution has become vanishingly small.
This shows that in a clean underdoped cuprate the regime of significant superconducting fluctuations
does not extend in temperature very far beyond $T_c$.
More quantitatively, the superconducting contribution to the Nernst coefficient $\nu/T$ in YBCO drops to 
0.1\% of its peak value at $T_c$ by $T \simeq 1.35~T_c$.

In Fig.~3, we see that the normal-state $\nu/T$ at 28 T is independent of 
temperature below $\sim 25$~K. 
Its large negative value at $T \to 0$ is completely and unambiguously due to quasi-particles. 
In Fig.~4, we compare the normal-state $\nu_a$ and $\nu_b$ as a function of temperature. 
We see that the large anisotropy characteristic of the pseudogap phase disappears below $\sim 25$~K.

The in-plane anisotropy of the normal-state $\nu(T)$ is plotted in Fig.~5, as the difference
$\nu_a/T - \nu_b/T$ (top panel) and the normalized difference $(\nu_a - \nu_b)/(\nu_a + \nu_b)$ (bottom panel).
The Nernst anisotropy is seen to rise just below the pseudogap temperature $T^\star$, defined as the temperature below which
the $a$-axis resistivity drops below its linear dependence at high temperature.\cite{Daou2010}
Upon cooling, it continues to rise, until it reaches a maximal value of 
$(\nu_a - \nu_b)/(\nu_a + \nu_b) \simeq 0.75$ 
({\it i.e.} $\nu_a/\nu_b \simeq 7$)
at $\sim 80$~K.
Upon further cooling, however, we now find that the anisotropy drops rapidly, with 
$(\nu_a - \nu_b)/(\nu_a + \nu_b) \to 0$ 
({\it i.e.} $\nu_a/\nu_b \to 1$)
as $T \to 0$.

%%%%%%%%%%%%%%%%%%%%%%%%%%%%%%%%%%%%%%%%%%%%%%%%%%%%%%%%%%%%%%%%%%%%%%%%%%
%%%%%%%%%%%%%%%%%%%%%%   DISCUSSION   %%%%%%%%%%%%%%%%%%%%%%%%%%%%%%%%%%%%
%%%%%%%%%%%%%%%%%%%%%%%%%%%%%%%%%%%%%%%%%%%%%%%%%%%%%%%%%%%%%%%%%%%%%%%%%%

\section{Discussion}

To elucidate the cause of this dramatic drop in the Nernst anisotropy, we turn to the Hall coefficient $R_{\rm H}(T)$.
In Fig.~5, the normal-state Hall angle $\theta_{\rm H}$ is plotted as $\tan \theta_{\rm H}=\rho_{xy}/\rho_{xx}$,
the ratio of Hall to longitudinal resistivities. 
Upon cooling, we see that the drop in $\tan \theta_{\rm H}(T)$ to negative values tracks precisely the drop in Nernst anisotropy.
(Note that from the Onsager relation, $\sigma_{xy} = - \sigma_{yx}$, $R_{\rm H}$ is independent of current direction in the basal plane.\cite{Segawa2004})

%%%%%%%%%%%%%%%%%%%%%%%%%%%%%%%%%%%%%%%%%%%%%%%%
%        FSR  
%%%%%%%%%%%%%%%%%%%%%%%%%%%%%%%%%%%%%%%%%%%%%%%%

\subsection{Fermi-surface reconstruction}

\subsubsection{Electron pocket}

Soon after the discovery of quantum oscillations in YBCO,\cite{Doiron-Leyraud2007}
the fact that the oscillations were seen on top of a large background of negative Hall resistance 
led to the interpretation that the oscillations come from orbits around an electron-like Fermi pocket.\cite{LeBoeuf2007}
This interpretation was later confirmed by the observation of a large negative Seebeck coefficient at low temperature.\cite{Chang2010}
Clinching evidence came recently from the quantitative agreement between the measured (negative) value of $S/T$ at $T \to 0$, on the one hand, 
and the magnitude of $S/T$ expected from the Fermi energy inscribed in the quantum oscillations, 
on the other hand, both obtained in YBCO at the same doping, namely $p=0.11$.\cite{Laliberte2011}
Therefore, in the doping interval from $p = 0.083$ to at least $p = 0.152$,\cite{LeBoeuf2011,Laliberte2011} 
the Fermi surface of YBCO in its non-superconducting ground state contains a small closed electron pocket. 
This pocket dominates the transport properties at low temperature, as discussed in detail in Ref.~\onlinecite{LeBoeuf2011}.
In particular, it produces a large (quasiparticle) Nernst signal. 
Applying Eq.~1 to YBCO at $p=0.11$, where quantum oscillations give $T_{\rm F} = 410 \pm 20$~K 
and $\mu = 0.02 \pm 0.006$~T$^{-1}$,\cite{Jaudet2008}
yields $|\nu/T| = 13 \pm 3$~nV/K$^2$T,
while the measured value at $p=0.11$ is
$\nu/T = -~13 \pm 3$~nV/K$^2$T,\cite{Laliberte2011}
in perfect agreement.
The somewhat smaller value at $p=0.12$, namely 
$\nu/T \simeq -~7$~nV/K$^2$T as $T \to 0$ (Fig.~4),
is probably due to the lower mobility expected of samples in the ortho-VIII phase (with $y=6.67$)
compared to those in the ortho-II phase (with $y = 6.54$), consistent with the much weaker 
quantum oscillations in the former.

The fact that $\nu_a \simeq \nu_b$ as $T \to 0$ shows that the electron pocket yields
transport properties that are isotropic in the plane. 
This explains two features of the transport in YBCO.
The first is the jump in the in-plane anisotropy of the resistivity as the doping drops below $p=0.08$.\cite{Ando2002,Sun2004}
Indeed, it was recently discovered that the electron pocket disappears suddenly as the doping is reduced below a critical
value $p = 0.08$,\cite{LeBoeuf2011}
in the sense that for $p < 0.08$ both Hall\cite{LeBoeuf2011} and Seebeck\cite{Laliberte2011} coefficients depend weakly on temperature
and remain positive at $T \to 0$. 
This change in Fermi-surface topology (or Lifshitz transition) coincides with a ten-fold increase in resistivity at $T \to 0$,\cite{LeBoeuf2011}
showing that the high-conductivity part of the Fermi surface has disappeared.
Once the high-mobility electron pocket is removed, the in-plane anisotropy 
ratio $\rho_a/\rho_b$ rises (see Ref.~\onlinecite{LeBoeuf2011}).

The second feature is the rapid loss of in-plane anisotropy in the Nernst coefficient upon cooling.
As shown in Fig.~5, the precipitous drop in the anisotropy ratio $(\nu_a - \nu_b)/(\nu_a + \nu_b)$ below 80~K tracks closely the fall in the 
Hall signal (plotted as $\tan \theta_{\rm H}$) towards large negative values.
As the electron pocket becomes increasingly dominant upon cooling, {\it i.e.} as its mobility $\mu \propto \tan \theta_{\rm H}$ increases,
the Nernst signal becomes increasingly isotropic.

\subsubsection{Stripe order}

The natural explanation for the emergence of an electron pocket in a hole-doped cuprate is the onset of a new periodicity that breaks
the translational symmetry of the lattice and thus reduces the Brillouin zone, causing a reconstruction of the large hole Fermi surface
into smaller pieces.\cite{Chakravarty2008}
A recent study that compares YBCO to the hole-doped cuprate Eu-LSCO found that the Seebeck coefficient behaves in essentially 
identical fashion in the two materials, as a function of both temperature and doping:\cite{Laliberte2011} 
$S/T$ drops to negative values (of very similar magnitude) below the same peak temperature, 
the sign-change temperature $T_0^S$ is maximal at $p = 1/8$ in both cases, 
the drop in $S/T$ disappears below the same critical doping $p = 0.08$.
So the same Fermi-surface reconstruction must be taking place in Eu-LSCO as in YBCO, 
pointing to a generic mechanism of hole-doped cuprates.

Now in Eu-LSCO, charge modulations are observed by x-ray diffraction 
at low temperature,\cite{Fink2010}
over the entire doping range where $S/T < 0$.\cite{Laliberte2011}
Spin modulations are most likely also present, as observed in the closely related material
La$_{1.6-x}$Sr$_x$Nd$_{0.4}$CuO$_4$ (Nd-LSCO).\cite{Ichikawa2000}
Called `stripe order', these spin and charge modulations break the translational symmetry of the CuO$_2$ planes, 
and so will cause a reconstruction of the Fermi surface. 
Calculations for stripe order at $p = 1/8$ show that an electron pocket will generically appear,\cite{Millis2007}
causing the quasiparticle Nernst signal to be strongly enhanced.\cite{Hackl2010}
It is then reasonable to infer that stripe order causes the Fermi-surface reconstruction in these 
underdoped cuprates.

Because stripe order involves unidirectional spin and/or charge modulations, it also breaks the four-fold rotational 
symmetry of the CuO$_2$ planes.\cite{Kivelson2003,Vojta2009}
So the reconstructed Fermi surface is expected to have strong in-plane anisotropy, 
manifest in the calculations by the presence of quasi-1D open sheets.\cite{Millis2007}
However, if the conductivity of the relatively isotropic electron pockets is much higher, at low temperature, than that of these open sheets, the 
inherent anisotropy of the latter will only show up in transport when the electron pocket disappears, as it does below $p=0.08$.

\subsection{The pseudogap phase}

We have focused so far on the non-superconducting ground state at $T \to 0$, with its stripe order and 
reconstructed Fermi surface.
Let us now ask what happens when the temperature is raised.

The presence of the electron pocket persists at least as long as the Hall coefficient is negative.
In YBCO at $p=0.12$, $R_{\rm H}(T)$ changes from negative to positive at the sign-change
temperature $T_0^H = 70$~K, above $T_c = 66$~K.\cite{LeBoeuf2007,LeBoeuf2011}
Of course, the drop in $R_{\rm H}(T)$ starts at higher temperature, namely at the peak in $R_{\rm H}(T)$ near 90~K (see Fig.~5). 
In fact, the onset of the downturn
is really at the temperature $T_{\rm H}$ where the curvature changes from upward at high temperature to downward at low temperature, 
{\it i.e.} at the inflexion point.
At $p = 0.12$, $T_{\rm H} \simeq 120$~K (see Fig.~5). 
In Fig.~1, this temperature $T_{\rm H}$ is plotted vs $p$ on the phase diagram of YBCO.
We see that it lies inside the pseudogap phase, between the crossover temperature $T^\star$ and the zero-field
superconducting temperature $T_c$.
This means that the electron pocket starts emerging at temperatures well above the onset of superconductivity,
and it does so even in small magnetic fields. In this sense, the onset of Fermi-surface reconstruction is not field-induced;
it is a property of the zero-field pseudogap phase.    

If $T_{\rm H}$ marks the onset of Fermi-surface reconstruction in YBCO as detected in the Hall effect,
what corresponding characteristic temperature do we obtain from other transport properties?
From the Seebeck coefficient $S/T$ vs $T$, a similar characteristic temperature is obtained, with 
$T_{\rm S} \simeq 100$~K at $p = 0.12$.\cite{Chang2010,Laliberte2011}
However, the Nernst coefficient, plotted as $\nu/T$ vs $T$, starts its drop to large negative values at a temperature $T_{\nu}$ 
which is significantly higher, namely $T_{\nu} \simeq 225$~K at $p=0.12$.\cite{Daou2010}
($T_{\nu}$ is independent of direction, the same whether it is measured in $\nu_a$ or $\nu_b$.\cite{Daou2010})
Calculations show that the quasiparticle Nernst effect is an extremely sensitive probe of Fermi-surface distortions such as would 
arise from broken rotational symmetry.\cite{Hackl2009}
The value of $T_{\nu}$ is plotted as a function of doping in the phase diagram of Fig.~1.
We see that $T_{\nu} \simeq 2~T_{\rm H}$.
Now $T_{\nu}$ coincides with the temperature $T_{\rho}$ below which the in-plane ($a$-axis) resistivity $\rho_a(T)$ of YBCO deviates from
its linear temperature dependence at high temperature. 
This $T_{\rho}$ is regarded as the standard definition of the pseudogap crossover temperature $T^\star$.\cite{Ito1993}
The fact that $T_{\nu} = T_{\rho}$ at all dopings shows that the drop in $\nu/T$ to negative values is a property of the pseudogap phase.

Given that the large value of $\nu/T$ at $T \to 0$ is firmly associated with the small high-mobility electron pocket in YBCO, 
can $T_{\nu}$ therefore be regarded as the onset of Fermi-surface reconstruction as detected in the Nernst effect?
By the same token, can $T_{\rho}$ be regarded as the onset of incipient Fermi-surface reconstruction detected in the resisitivity?
If so, then the pseudogap phase would be the high-temperature precursor of the stripe-ordered phase present at low temperature.

One evidence in support of a stripe-precursor scenario is the fact that the enhancement of the Nernst coefficient $\nu/T$ below $T^\star$ is anisotropic,
that it breaks the rotational symmetry of the CuO$_2$ planes. This, of course, is a characteristic signature of stripe ordering.
Indeed, the sequence of broken symmetries, first rotational then translational, is expected in the gradual process of stripe ordering.\cite{Kivelson1998}
The sequence is called `nematic to smectic' ordering.
The fact that $T_{\nu}$ and $T_{\rho}$ are higher than $T_{\rm S}$ and $T_{\rm H}$, by roughly a factor 2, may come from the role of scattering in the various transport coefficients. Indeed, while $\nu$ and $\rho$ both depend directly on the scattering rate (or mobility $\mu$), 
with $\nu \propto \mu$ and $\rho \propto 1/\mu$, $S$ and $R_{\rm H}$ do not (at least in a single band model).  
In other words, if stripe fluctuations affect the transport primarily through the scattering rate, then we would expect $\rho$ and $\nu$ to be sensitive 
to the onset of stripe fluctuations, but not $S$ and $R_{\rm H}$.
As we shall now see, similar precursor effects are observed in the iron-pnictide superconductors.

%%%%%%%%%%%%%%%%%%%%%%%%%%%%%%%%%%%%%%%%%%%%%%%%
%    PNCITIDES  
%%%%%%%%%%%%%%%%%%%%%%%%%%%%%%%%%%%%%%%%%%%%%%%%

\subsection{Comparison with pnictide superconductors}

It is instructive to compare the cuprate superconductor YBCO with
the iron-pnictide superconductor Ba(Fe$_{1-x}$Co$_x$)$_2$As$_2$ (Co-Ba122).
In the parent compound BaFe$_2$As$_2$, a well-defined antiferromagnetic order sets in 
below a critical temperature $T_{\rm N} = 140$~K.\cite{Canfield2010}
This order is unidirectional, with chains of ferromagnetically aligned spins along the $b$-axis of the orthorhombic 
crystal structure alternating antiferromagnetically in the perpendicular direction ($a$-axis).
In other words, this is a form of `spin-stripe' order, which breaks both the translational and rotational symmetry of the original
tetragonal lattice (present well above $T_{\rm N}$).
As Co is introduced, $T_{\rm N}$ falls, and superconductivity appears, with $T_c$ peaking at the point where it crosses $T_{\rm N}$.
In other words, in the underdoped region the normal state is characterized by spin-stripe order for some range of temperature above $T_c$.
This order causes a reconstruction of the Fermi surface, which leads to a change in the in-plane resistivity:
a drop below $T_{\rm N}$ at low Co concentration $x$, an upturn at intermediate $x$.
However, the upturn starts well above $T_{\rm N}$.
One can define a temperature $T_\rho$ below which the roughly linear $T$ dependence at high temperature turns upwards.
For $3 \% < x < 5 \%$, $T_\rho \simeq 2$~$T_{\rm N}$.\cite{FisherSCIENCE2010}
Moreover, the rise in $\rho$ is anisotropic: a strong in-plane anisotropy appears at $T_\rho$.\cite{FisherSCIENCE2010}
At higher doping, above the quantum critical point where antiferromagnetic order disappears (in the absence of superconductivity),
both the upturn and the anisotropy in $\rho$ vanish.\cite{FisherSCIENCE2010}
 
The overall phenomenology is seen to be remarkably similar to that of YBCO.
The unidirectional order that breaks translational symmetry and reconstructs the Fermi surface at low temperature is preceded at high temperature
by a regime with strong in-plane transport anisotropy, evidence of broken rotational symmetry. 
In Co-Ba122, it is very natural to view this regime as the nematic precursor to the smectic phase 
at low temperature. 
The analogy supports the view that the pseudogap phase in YBCO is just such a precursor to stripe order.

It is interesting to note that in Co-Ba122 the in-plane anisotropy in $\rho$ is not largest at $x=0$, where the order is strongest ($T_{\rm N}$ is highest).
Indeed, the ratio $\rho_b/\rho_a$ at $T \to 0$ is larger at $x > 2~\%$ than at $x=0$.\cite{FisherSCIENCE2010}
This could well be due to a short-circuiting effect similar to that observed in YBCO, whereby a small and isotropic Fermi pocket of high mobility dominates the 
conductivity. 
Indeed, the concentration $x \simeq 0.02$ in Co-Ba122 appears to be a Lifshitz critical point, as suggested 
by ARPES experiments that 
reveal the existence of a small hole pocket for $x<0.02$, and not above.\cite{KaminskiNATPHYS2010}
%
%This is consistent with quantum oscillation measurements
%that observe three small pockets in the parent compound 
%BaFe$_2$As$_2$.\cite{AnalytisPRB2009}
%
%Hall measurements also suggest an important change in the band 
%structure for $x<0.02$,\cite{AlloulPRL2009}
%where $R_{\rm H}$ is large and negative.
%is large and negative for $x<0.02$ and take much smaller values for 
%$x>0.02$. 
%Hence the combination of transport and ARPES measurements on Co-Ba122 point to the 
%existence of small high-mobility Fermi pocket for $x<0.02$.
%
Such a pocket could short-circuit the in-plane anisotropy coming from other parts of the Fermi surface.

%%%%%%%%%%%%%%%%%%%%%%%%%%%%%%%%%%%%%%%%%%%%%%%%%%%%%%%%%%%%%%%%%%%%%
%%%%%%%%%%%%%%%%%%%%%   CONCLUSION  %%%%%%%%%%%%%%%%%%%%%%%%%%%%%%%%%
%%%%%%%%%%%%%%%%%%%%%%%%%%%%%%%%%%%%%%%%%%%%%%%%%%%%%%%%%%%%%%%%%%%%%

\section{Conclusion}

In summary, the Nernst effect is a highly sensitive probe of electronic transformations in
metals. In the cuprate superconductor YBCO, the onset of the pseudogap phase
at $T^\star$ causes a 100-fold enhancement of the quasiparticle Nernst coefficient, 
which goes smoothly from 
$\nu_b/T = +~0.07$~nV/K$^2$T at $T^\star$
to 
$\nu_b/T = -~7$~nV/K$^2$T at $T \to 0$, in the normal (non-superconducting) state.
This enhancement is strongly anisotropic in the plane, showing that the pseudogap phase 
breaks the rotational symmetry of the CuO$_2$ planes.
When the Fermi-surface reconstructs at low temperature, the formation of a high-mobility
electron pocket short-circuits this in-plane anisotropy. 
The magnitude of $\nu/T$ at $T \to 0$ is in perfect agreement with the value
expected from the small closed Fermi pocket detected in quantum oscillations.
The negative sign proves that the signal comes from quasiparticles and 
not superconducting fluctuations or vortices.
There is compelling evidence that the Fermi-surface reconstruction is caused by
a stripe order which breaks the translational symmetry of the CuO$_2$ planes at low temperature,\cite{Laliberte2011}
but also its rotational symmetry.
The sequence of broken symmetries upon cooling, first rotational then translational,
suggests a process of nematic-to-smectic stripe ordering, similar to that observed in 
the iron-pnictide superconductor Co-Ba122, where a phase of spin-stripe antiferromagnetic order prevails
in the underdoped region of the temperature-concentration phase diagram.
The analogy suggests that the enigmatic pseudogap phase of hole-doped cuprates is also a high-temperature precursor
of a stripe-ordered phase, with unidirectional charge and/or spin modulations.
This nematic interpretation is consistent with the in-plane anisotropy of the spin fluctuation spectrum
detected by neutron scattering in underdoped YBCO.\cite{Hinkov2008}

%%%%%%%%%%%%%%%%%%%%%%%%%%%%%%%%%%%%%%%%%%%%%%%%%%%%%%%%%%%%%%%%%%%%%
%%%%%%%%%%%%%%%%%%%%%%   ACKNOWLEDGEMENTS   %%%%%%%%%%%%%%%%%%%%%%%%%
%%%%%%%%%%%%%%%%%%%%%%%%%%%%%%%%%%%%%%%%%%%%%%%%%%%%%%%%%%%%%%%%%%%%%

\section{Acknowledgements}

We thank J. Corbin and J. Flouquet for assistance with measurements in Sherbrooke and at the LNCMI in Grenoble,
respectively.
J.C. was supported by Fellowships from the Swiss SNF and 
FQRNT.
Part of this work was supported by Euromagnet under the EU contract RII3-CT-2004-506239.
C.P. and K.B. acknowledge support from the ANR project DELICE.
L.T. acknowledges support from the Canadian Institute for
Advanced Research, a Canada Research Chair, NSERC, CFI and
FQRNT.

%%%%%%%%%%%%%%%%%%%%%%   REFERENCES

\end{document}